\documentclass[aps,pra,onecolumn,superscriptaddress,notitlepage,nofootinbib,longbibliography, colorlinks=true]{revtex4-2}
\usepackage{mathrsfs,graphicx,amsfonts,amssymb,amsmath}
\usepackage[figuresright]{rotating}
\usepackage[dvipsnames]{xcolor}
\usepackage[detect-all=true,group-minimum-digits=4]{siunitx}
\usepackage{xspace}

\definecolor{mypink3}{cmyk}{0, 0.7808, 0.4429, 0.1412}
\definecolor{mypink1}{rgb}{0.858, 0.188, 0.478}
\definecolor{mypink2}{RGB}{219, 48, 122}
\usepackage[colorlinks=true, allcolors=teal]{hyperref}

\definecolor{lapislazuli}{rgb}{0, 0, 1}
\definecolor{YKblue}{rgb}{0.0, 0.18, 0.65}
\definecolor{carmine}{rgb}{0.81, 0.09, 0.03}
\definecolor{lavender}{rgb}{0.84, 0.49, 0.87}
\usepackage{nicefrac,xfrac}
\usepackage[normalem]{ulem}

\usepackage[fleqn]{mathtools}

\begin{document}

\title{Quantum correlations enhanced in hybrid optomechanical system via phase tuning}

\author{K. B. Emale}
\email{emale.kongkui@facsciences-uy1.cm}
\affiliation{Department of Physics, Faculty of Science, University of Yaounde I, P.O.Box 812, Yaounde, Cameroon}

\author{J.-X. Peng}
\affiliation{School of Physics and Technology, Nantong University, Nantong, 226019, People’s Republic of China}

\author{P. Djorwé}
\email{djorwepp@gmail.com}
\affiliation{Department of Physics, Faculty of Science, 
University of Ngaoundere, P.O. Box 454, Ngaoundere, Cameroon}
\affiliation{Stellenbosch Institute for Advanced Study (STIAS), Wallenberg Research Centre at Stellenbosch University, Stellenbosch 7600, South Africa}

\author{A. K. Sarma}
\affiliation{Department of Physics, Indian Institute of Technology Guwahati, Guwahati 781039, India}

\author{Abdourahimi}
\affiliation{Department of Physics, Faculty of Science, University of Yaounde I, P.O.Box 812, Yaounde, Cameroon}

\author{A.-H. Abdel-Aty}
\affiliation{Department of Physics, College of Sciences, University of Bisha, Bisha 61922, Saudi Arabia}

\author{K.S. Nisar}
\affiliation{Department of Mathematics, College of Science and Humanities in Alkharj, Prince Sattam Bin Abdulaziz University, Alkharj 11942, Saudi Arabia}

\author{S. G. N. Engo}
\affiliation{Department of Physics, Faculty of Science, University of Yaounde I, P.O.Box 812, Yaounde, Cameroon}

\begin{abstract}
This work presents a theoretical framework for enhancing quantum correlations in a hybrid double-cavity optomechanical system that hosts an atomic ensemble. We investigate the role of the coupling phase $\phi$ between cavity 1 and the atomic ensemble in optimizing quantum correlations, i.e., bipartite/tripartite quantum entanglement and quantum discord. By employing metrics such as logarithmic negativity for bipartite entanglement and minimum residual contangle for genuine tripartite entanglement, we demonstrate that tuning the phase $\phi$ is essential for maximizing photon-phonon entanglement. Specifically, we find that optimal entanglement occurs at $\phi=n\pi$, with distinct conditions for odd and even integers $n$. Our results also indicate that the quantum entanglement achieved in this system is robust against thermal fluctuations, making it a promising candidate for applications in quantum information processing and quantum computing. Furthermore, this research highlights the significance of phase tuning in controlling quantum correlations, paving the way for advancements in quantum technologies.
\end{abstract}

\maketitle

\section{Introduction}

Quantum correlations is an interesting property of quantum physics, distinguished by the nonseparability of quantum states over considerable distances. This distinctive aspect has been instrumental in driving notable progress in quantum technologies, encompassing quantum information processing \cite{ekert1991}, quantum teleportation, quantum cryptography \cite{bennett1993}, sensing technologies \cite{tao2024,Djorwe2019}, and metrology \cite{dooley2023}. While research has largely concentrated on microscopic systems—such as atoms and photons—there is an increasing interest in macroscopic systems, such as mechanical oscillators, due to their potential to establish stable quantum correlations \cite{ockeloen2018}.
Recently, entanglement in particular has been effectively prepared and manipulated in various microscopic systems both theoretically and experimentally. These systems are atoms~\cite{Sun:15}, photons~\cite{PhysRevA.79.042336}, and ions~\cite{PhysRevLett.74.4091}. The successful generation and control of entanglement in these microscopic systems has not only contributed to the advancement of research work but also facilitates potential applications of entanglement in quantum information processing and quantum computing. However, entanglement of atoms with macroscopic systems has also garnered significant interest~\cite{KUSSIA2024129920}.

Cavity optomechanical (COM) systems have become prominent platforms for the investigation of quantum phenomena. By employing radiation pressure to couple mechanical oscillators with optical fields, COM systems enable the generation of entanglement and the establishment of other quantum correlations \cite{mancini2002,Sarma2021,Agasti2024}. These quantum correlations are interesting for the improvement of quantum technologies, and may be used to enhance sensors sensitivity \cite{Tchounda_2023,Phil.2024}. Nonetheless, conventional radiation pressure mechanisms frequently fail to achieve significant quantum effects. Recent theoretical developments have introduced methods to enhance radiation pressure by as much as six orders of magnitude, thus facilitating entry into a regime of strong coupling \cite{Pirkkalaine2014}.
Whereas bipartite entanglement within COM systems has been the subject of extensive investigation \cite{Djorwe2014,Bougouffa2020,Rostand2024,Tesfaye2020,Tadesse2024,Tchodimou2017}, tripartite entanglement remains comparatively under-explored \cite{Hmouch2023,Wen2024}. Moreover, the task of enhancing and maintaining entanglement in the face of environmental decoherence—most notably at elevated temperatures—constitutes a significant challenge \cite{genes2008}.  In addition to entanglement, quantum discord is recognized as a vital measure of quantumness in bipartite states, including both entangled and separable states \cite{Ollivier2001}. In continuous variable systems, Gaussian quantum discord (GQD) serves to quantify correlations that extend beyond mere entanglement \cite{Giorda2010}. Despite its significance, GQD is frequently disregarded in favor of more conventional measures of quantification.

The primary objective of this research is to address these challenges by investigating a hybrid system that incorporate multiple degrees of freedom to enhance quantum correlations. In particular, we propose a theoretical framework aiming to achieves robust quantum correlations within a hybrid optomechanical system comprising a double cavity in conjunction with an atomic ensemble. Our approach utilizes logarithmic negativity to evaluate bipartite entanglement, minimum residual contangle for genuine tripartite entanglement, and Gaussian quantum discord (GQD) to assess correlations that transcend mere entanglement \cite{Vidal2002}. Importantly, we demonstrate that manipulating the coupling phase between the cavities and the atomic ensemble is vital for maximizing quantum correlations. Our findings indicate that optimal bipartite entanglement is realized at specific phase values, which under certain conditions, facilitate genuine tripartite entanglement. This study presents several novel contributions: (1) it establishes a conceptual framework for the analysis of quantum correlations in hybrid optomechanical systems; (2) it identifies phase tuning as a pivotal element for the enhancement of tripartite entanglement; and (3) it highlights the robustness of quantum correlations in the presence of thermal fluctuations. These insights pave the way for advanced quantum information technologies and applications in quantum computing.

The organization of this paper is delineated as follows. In \autoref{sec:II} we provide a detailed exposition of the hybrid double-cavity optomechanical system that incorporates an atomic ensemble, accompanied by the formulation of the Hamiltonian and the derivation of the equations of motion. Furthermore, the output field of the system is calculated.  Numerical analysis for quantifying bipartite and tripartite entanglement, as well as the Gaussian quantum discord (GQD) are carrying out in \autoref{sec:III}. Finally, our conclusions are encapsulated in \autoref{sec:IV}.

\section{Model and dynamics}\label{sec:II}

\subsection{Hamiltonian of the system}\label{AII}

\begin{figure}[htp!]
\setlength{\lineskip}{0pt}
\centering
\includegraphics[width=.5\linewidth]{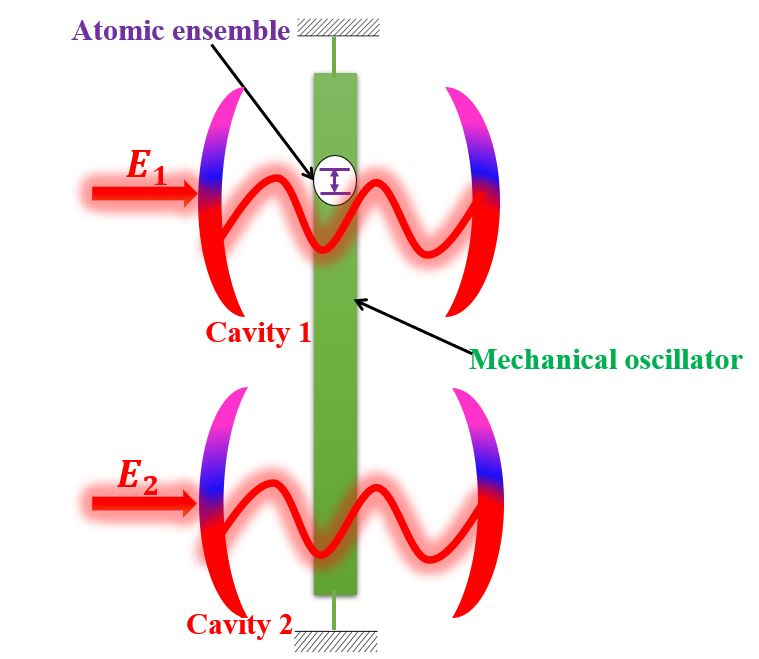}
\caption{A hybrid double--cavity optomechanical system, where  atomic ensemble is trapped in  cavity 1, and trapped atomic ensemble couple to a mechanical oscillator. The coupling fields with amplitudes $E_1$ and $E_2$ are used to drive the two cavities.}
 \label{fig:1}
\end{figure}

Our benchmark system as referred in \autoref{fig:1} is an integrated hybrid atomic ensemble-cavity optomechanical configuration consisting of two optical cavities. The first cavity hosts a trapped atomic ensemble interacting with a mechanical oscillator. Both cavities are coupled to the mechanical oscillator. The complete Hamiltonian governing this system is ($\hbar=1$),
\begin{equation}
\mathcal{H}=\mathcal{H}_o+\mathcal{H}_{at}+\mathcal{H}_{in}+\mathcal{H}_{dr},
\end{equation}
where   
\begin{equation}
\mathcal{H}_o=\Delta_1c_1^\dagger c_1+\Delta_2c_2^\dagger c_2+\omega_mb^\dagger b,
\end{equation}
defines the free Hamiltonian corresponding to the cavity fields and the mechanical oscillator, with  $c_j(c_j^{\dagger})$ and $b(b^\dagger)$, the annihilation (creation) operators of the cavity and mechanical modes, respectively. The detuning of the cavity fields is defined as $\Delta_j=\omega_{cj}-\omega_{l}$ and the mechanical mode frequency is denoted by $\omega_m$. The atomic ensemble Hamiltonian $\mathcal{H}_{at}$ is given by,
\begin{equation}
\begin{aligned}
\mathcal{H}_{at}&=\frac{\Delta_{at}}{2}\sum_{j=1}^{N}\sigma^z_j+\left(J^{\ast}_{ac}c_1\sum_{j=1}^{N}\sigma^{+}_j+J_{ac}c^{\dagger}_1\sum_{j=1}^{N}\sigma^{- }_j\right)+J_{ab}\left(\sum_{j=1}^{N}\sigma^{+}_j+\sum_{j=1}^{N}\sigma^{-}_j\right)(b+b^\dagger),
\end{aligned}
\end{equation}
where $J_{ac}$ is the coupling strength of the cavity mode 1 with the atomic ensemble and $J_{ab}$ is the coupling strength of the atomic ensemble with the mechanical mode. The spin-1/2 of the Pauli operators $\sigma_j^z$ and $\sigma_j^{\pm}$ describe the $j^{th}$ two-level atom in the ensemble \cite{Momeni2018} which satisfies the commutation relations $\left[\sigma^{+}_j, \sigma^{-}_j\right]=\sigma_j^z$ and $\left[\sigma^z_j,\sigma^{\pm}_j\right]=\pm2\sigma_j^{\pm}$. The detuning of the atomic ensemble mode  relative to the driving field  is  $\Delta_{at}=\omega_{at}-\omega_{l}$. To carry out our numerical simulations, some  assumptions will be considered for simplicity.

To illustrate, the number of atoms in the excited state is significantly smaller than the total number of atoms, indicating that the atomic ensemble is in a low excitation state. In light of this assumption, the atomic ensemble can be regarded as a bosonic mode.  Therefore, we define the creation and annihilation operators of this mode as $a$ and $a^\dagger$, which are related to the Pauli matrices through the Holstein-Primakoff representation \cite{Momeni2018}. The collective operators of the atomic ensemble are then defined in accordance with the approaches outlined in \cite{Momeni2018,Nie2016},
\begin{equation*}
\frac{1}{N}\sum_{j=1}^{N}\sigma^-_j=a,~
\frac{1}{N}\sum_{j=1}^{N}\sigma^+_j=a^\dagger,~
\sum_{j=1}^{N}\sigma^z_j=2a^{\dagger}a-N,
\end{equation*}
where the operator $a$ obeys the commutation relation $[a,a^\dagger]=1$. From the above assumption, the atomic ensemble Hamiltonian yields,    
\begin{equation}
\mathcal{H}_{at}=\Delta_{at}a^{\dagger}a+\left(J^{\ast}_{ac}c_1a^\dagger+J_{ac}c^{\dagger}_1a\right)+J_{ab}\left(a+a^\dagger\right)\left(b+b^\dagger\right).
\end{equation}
The Tavis-Cummings type of coupling is employed between the cavity mode and the atomic ensemble, thereby enabling the capture of the phase relationship between the atoms and the cavity field. The phase of the coupling can be introduced by considering the coupling strength, $J_{ac}$, to be complex. It is of paramount importance to tune this phase in order to elucidate the manner in which the atoms interact with the cavity field. Conversely, the interaction between the mechanical oscillator and atomic ensemble is of the Dicke type, thereby enabling robust coupling between the mechanical oscillator and the atomic ensemble. The strong coupling between a mechanical oscillator and a single atom was achieved in Ref.~\cite{Hammerer2009}. The Hamiltonian that defines the interaction between the cavity modes and the mechanical mode can be expressed as,
\begin{equation}
\mathcal{H}_{in}=g_1c_1^\dagger c_1(b^\dagger+b)+g_2c_2^\dagger c_2(b^\dagger+b),
\end{equation} 
where $g_j$ ($j=1,2$) is the coupling strength between the cavity $j$ and mechanical mode. The final form of the Hamiltonian that captures the interaction between the cavity modes and the coupling fields is,
\begin{equation}
\mathcal{H}_{dr}=i(E_1c_1^\dagger-E_1^\ast c_1)+i(E_2c_2^\dagger-E_2^\ast c_2),
\end{equation}
where $E_1$ and $E_2$ are the driving amplitudes, with $E_j$ related to the laser power $\mathcal{P}_j$ as, $E_j=\sqrt{\frac{2\mathcal{P}_j\kappa_j}{\hbar\omega_{l}}}$, in which $\kappa_j$ is the decay rate of the cavity $j$.

\subsection{Quantum Langevin Equations} \label{IIB}

The dynamics of our system is captured through the Quantum Langevin Equations (QLEs), which are derived from the aforementioned Hamiltonian, by applying the Heisenberg equations of motion. Therefore, the QLEs of our system yield,
\begin{equation}\label{eq:6}
\begin{aligned}
\dot{c_1}&=-\left(i\Delta_1+\kappa_1\right)c_1-ig_1c_1(b+b^\dagger)-iJ_{ac}a+E_1 +\sqrt{2\kappa_1}c_1^{in},\\
\dot{c_2}&=-\left(i\Delta_2+\kappa_2\right)c_2-ig_2c_2(b+b^\dagger)+\mathcal{E}_2+\sqrt{2\kappa_2}c_2^{in},\\
\dot{a}&=-\left(i\Delta_{at}+f\right)a-iJ^{\ast}_{ac}c_1-iJ_{ab}(b+b^\dagger)+\sqrt{2f}a^{in},\\
\dot{b}&=-\left(i\omega_m+\gamma_m\right)b-ig_1c^{\dagger}_1c_1-ig_2c^{\dagger}_2c_2-iJ_{ab}(a+a^\dagger)+\sqrt{2\gamma_m}b^{in},
\end{aligned}
\end{equation}
where we have taken into account the input noise and dissipation of each mode. To simplify the analysis, we linearize the QLEs around the steady-state values of the system's operators. This linearization is valid under the assumption that the system operates in the regime of strong optical driving, where the quantum fluctuations around the steady-state values are small. Mathematically, each operator $\mathcal{O}$ is expanded as, $\mathcal{O}=\langle\mathcal{O}\rangle+\delta\mathcal{O}$, where $\mathcal{O}\equiv c_1,c_2,a,b$, $\langle\mathcal{O}\rangle\equiv\alpha_1, \alpha_2, \xi, \beta$ represents the steady-state mean value, and $\delta\mathcal{O}$ corresponds to the small quantum fluctuation around this mean value. The steady-state values are found by solving the classical (mean-field) equations of motion, and the fluctuations $\delta\mathcal{O}$ satisfy the linearized QLEs. This procedure allows us to reduce the complexity of the problem by focusing on the dynamics of the small fluctuations, which are assumed to evolve linearly, thereby making the analysis tractable. Importantly, this linearization is valid under the condition that the system remains stable, and the fluctuations are sufficiently small compared to the steady-state values. From the linearization process, one gets the following steady-state dynamical set of equations, 
\begin{equation}
\begin{aligned}
 \alpha_1&=\frac{E_1-iJ_{ac}\xi}{\left(i\Delta^\prime_1+\kappa_1\right)},\\
 \alpha_2&=\frac{E_2}{\left(i\Delta^\prime_2+\kappa_2\right)},\\
\xi&=-\frac{iJ^\ast_{ac}\alpha_1+2iJ_{ab}\text{Re}(\beta)}{\left(i\Delta_{at}+f\right)},\\
\beta&=-\frac{ig_1|\alpha_1|^2+ig_2|\alpha_2|^2+2iJ_{ab}\text{Re}(\xi)}{\left(i\omega_m+\gamma_m\right)},
\end{aligned}
\end{equation}
where $\Delta_j^\prime=\Delta_j+2g_j\text{Re}(\beta)$ is the effective detunings of the $j^{th}$ cavity. By assuming that $J_{ac}$ is complex, it can be written in the polar form as $J_{ac}=|J_{ac}|e^{i\phi}$, with $\phi$ representing the phase associated to the coupling strength of the atomic ensemble and the first cavity. The dynamics of fluctuations derived from the linearization yields,
\begin{equation}\label{eq:LinEq}
\begin{aligned}
\delta\dot{c_1}&=-\left(i\Delta^\prime_1+\kappa_1\right)\delta c_1-i|J_{ac}|e^{i\phi}\delta a-i\mathcal{G}_1\left(\delta b+\delta b^\dagger\right)+\sqrt{2\kappa_1}c_1^{in},\\
\delta\dot{c_2}&=-\left(i\Delta^\prime_2+\kappa_2\right)\delta c_2-i\mathcal{G}_2\left(\delta b+\delta b^\dagger\right)+\sqrt{2\kappa_2}c_2^{in},\\
\delta\dot{a}&=-\left(i\Delta_{at}+f\right)\delta a-i|J_{ac}|e^{-i\phi}\delta c_1-iJ_{ab}(\delta b+\delta b^\dagger)+\sqrt{2f}a^{in},\\
\delta\dot{b}&=-\left(i\omega_m+\gamma_m\right)\delta b-i\mathcal{G}_1(\delta c_1+\delta c^\dagger_1)-i\mathcal{G}_2(\delta c_2+\delta c^\dagger_2)-iJ_{ab}(\delta a+\delta a^\dagger)+\sqrt{2\gamma_m}b^{in}.
\end{aligned}
\end{equation}
The effective optomechanical couplings are defined as $\mathcal{G}_1=g_1|\alpha_1|$ and $\mathcal{G}_2=g_2|\alpha_2|$. The quantum noise terms $c_j^{\text{in}}$, $a^{\text{in}}$, and $b^{\text{in}}$, corresponding to the input fields for the optical cavities, atomic ensemble, and mechanical oscillator respectively, are assumed to be Gaussian and Markovian. These noise terms arise due to the coupling of each mode to its thermal environment, and they obey the following correlation relation~\cite{Sarma2021},
\begin{equation}
\begin{aligned}
&\langle c_j^{\text{in}}\rangle=\langle a^{\text{in}}\rangle=\langle b^{\text{\text{in}}}\rangle=0,\\
&\langle c_j^{\text{in}}(t)c_j^{\text{in}\dagger}(t^\prime)\rangle=\delta(t-t^\prime),\\
&\langle a^{\text{in}}(t)a^{\text{in}\dagger}(t^\prime)\rangle=\delta(t-t^\prime),\\
&\langle b^{\text{in}}(t)b^{\text{in}\dagger}(t^\prime)\rangle=(n_{th}+1)\delta(t-t^\prime),\\
&\langle b^{\text{in}\dagger}(t)b^{\text{in}}(t^\prime)\rangle=n_{th}\delta(t-t^\prime),
\end{aligned}
\end{equation}
where $n_{\text{th}}=\left\{\exp\left(\frac{\hbar\omega_m}{k_B\text{T}}\right)-1\right\}^{-1}$ denotes the thermal phonon number at temperature $T$ and $k_B$ is the Boltzmann constant. These correlations ensure that the noise introduced by the thermal environment is accounted for, particularly for the mechanical mode, which is more sensitive to thermal noise.

It is noteworthy that the QLEs describe the interplay between the optical, mechanical, and atomic degrees of freedom in the system. The mechanical oscillator is driven by both the radiation pressure from the optical cavities and the interaction with the atomic ensemble, while the optical modes are influenced by the motion of the mechanical oscillator. The linearized QLEs reveal how small quantum fluctuations propagate through the system, giving rise to quantum correlations, including entanglement between the different modes. In particular, the effective optomechanical coupling terms $\mathcal{G}_1$ and $\mathcal{G}_2$, which appear in the set of \autoref{eq:LinEq}, depend on the steady-state intracavity fields and characterize the strength of the interaction between the optical and mechanical modes. By tuning these coupling terms, one can manipulate the dynamics of the system, allowing for the control of quantum correlations such as entanglement.

In order to fully characterize the quantum fluctuations of the optical, atomic, and mechanical modes, it is useful to define quadrature operators which allow us to express the system’s dynamics in terms of measurable quantities. Unlike the complex field operators, which may be less intuitive, the quadratures provide a clearer physical interpretation analogous to position and momentum in mechanical systems. Moreover, quadrature operators simplify the analysis of quantum correlations and entanglement, as they enable the formulation of the covariance matrix, which is essential for studying continuous variable quantum systems. This representation also facilitates the comparison of quantum fluctuations with classical noise, making the behavior of the system more transparent. We then define the quadratures operators for the optical, atomic modes and mechanical modes as,
\begin{equation}
\begin{aligned}
\delta x_j&=\frac{( \delta c_j+\delta c_j^{\dagger})}{\sqrt{2}}, \delta y_j=\frac{( \delta c_j- \delta c_j^{\dagger})}{i\sqrt{2}},\\
\delta q_{at}&=\frac{(\delta a+\delta a^{\dagger})}{\sqrt{2}},\delta p_{at}=\frac{(\delta a-\delta a^{\dagger})}{i\sqrt{2}},\\
\delta q&=\frac{(\delta b+\delta b^{\dagger})}{\sqrt{2}},\delta p=\frac{(\delta b-\delta b^{\dagger})}{i\sqrt{2}},
\end{aligned}
\end{equation}
where $j=1,2$. The corresponding input noises operators are,
\begin{equation}
\begin{aligned}
x_j^{\text{in}}&=\frac{(c_j^{\text{in}}+ c_j^{\text{\text{in}}\dagger})}{\sqrt{2}},~ y_j^{in}=\frac{( c_j^{\text{\text{in}}}- c_j^{in\dagger})}{i\sqrt{2}},\\
q^{\text{in}}_{at}&=\frac{( a^{\text{in}}+ a^{\text{in}\dagger})}{\sqrt{2}},~ p^{in}_{at}=\frac{( a^{\text{in}}- a^{in\dagger})}{i\sqrt{2}},\\
q^{\text{in}}&=\frac{( b^{in}+ b^{\text{in}\dagger})}{\sqrt{2}},~ p^{\text{in}}=\frac{( b^{\text{in}}- b^{\text{in}\dagger})}{i\sqrt{2}}.
\end{aligned}
\end{equation}
By rewriting the linearized equations in terms of field quadrature fluctuations, one obtains, 
\begin{equation}\label{eq:13}
\begin{aligned}
\delta\dot{x}_1=&\Delta^\prime_1\delta y_1-\kappa_1\delta x_1+|J_{ac}|\sin\phi\delta q_{at}+|J_{ac}|\cos\phi\delta p_{at}+\sqrt{2\kappa_1}x_1^{in},\\
\delta\dot{y}_1=&-\Delta^\prime_1\delta x_1-\kappa_1\delta y_1-|J_{ac}|\cos\phi\delta q_{at}+|J_{ac}|\sin\phi\delta p_{at}-2\mathcal{G}_1\delta q+\sqrt{2\kappa_1}y_1^{in},\\
\delta\dot{x}_2=&-\kappa_2\delta x_2+\Delta^\prime_2\delta y_2+\sqrt{2\kappa_2}x_2^{in},\\
\delta\dot{y}_2=&-\Delta^\prime_2\delta x_2-\kappa_2\delta y_2-2\mathcal{G}_2\delta q+\sqrt{2\kappa_2}y_2^{in},\\
\delta\dot{q}_{at}=&-|J_{ac}|\sin\phi\delta x_1+|J_{ac}|\cos\phi\delta y_1-f\delta q_{at}+\Delta_{at}\delta p_{at}+\sqrt{2f}q_{at}^{in},\\
\delta\dot{p}_{at}=&-|J_{ac}|\cos\phi\delta x_1-|J_{ac}|\sin\phi\delta y_1-\Delta_{at}\delta q_{at}-f\delta p_{at}-2J_{ab}\delta q+\sqrt{2f}p_{at}^{in},\\
\delta\dot{q}=&-\gamma_m\delta q+\omega_m\delta p+\sqrt{2\gamma_m}q^{in},\\
\delta\dot{p}=&-2\mathcal{G}_1\delta x_1-2\mathcal{G}_2\delta x_2-2J_{ab}\delta q_{at}-\omega_m\delta q-\gamma_m\delta p+\sqrt{2\gamma_m}p^{in}.
\end{aligned}
\end{equation}
\autoref{eq:13} can be written in a more compact form as,
\begin{equation}\label{eq:14}
\dot{\digamma}(t)=\mathcal{A}\digamma(t)+\mathcal{N}(t),
\end{equation}
where $\mathcal{A}$ is $8\times 8$ matrix which reads,
\begin{widetext}
\begin{equation}
\mathcal{A}=
\begin{pmatrix}
-\kappa_1&\Delta_1^\prime&0&0&|J_{ac}|\sin\phi&|J_{ac}|\cos\phi&0&0\\
-\Delta_1^\prime&-\kappa_1&0&0&-|J_{ac}|\cos\phi&|J_{ac}|\sin\phi&-2\mathcal{G}_1&0\\
0&0&-\kappa_2&\Delta_2^\prime&0&0&0&0\\
0&0&-\Delta_2^\prime&-\kappa_2&0&0&-2\mathcal{G}_2&0\\
-|J_{ac}|\sin\phi&|J_{ac}|\cos\phi&0&0&-f&\Delta_{at}&0&0\\
-|J_{ac}|\cos\phi&-|J_{ac}|\sin\phi&0&0&-\Delta_{at}&-f&-2J_{ab}&0\\
0&0&0&0&0&0&-\gamma_m&\omega_m\\
-2\mathcal{G}_1&0&-2\mathcal{G}_2&0&-2J_{ab}&0&-\omega_m&-\gamma_m\\
\end{pmatrix}.
\end{equation}
\end{widetext}
A formal solution of \autoref{eq:14} is given by,
\begin{equation}
\digamma(t)=\mathcal{M}(t)\digamma(0)+\int_{0}^{t}dt^{\prime}\mathcal{M}(t^{\prime})\mathcal{N}(t-t^{\prime}),
\end{equation}
where $\mathcal{M}(t)=\exp(\mathcal{A}t)$. It is noteworthy to mention that,
\begin{widetext}
\begin{equation}
\begin{aligned}
&\digamma^{T}(t)=\left(\delta x_1(t),\delta y_1(t),\delta x_2(t),\delta y_2(t),\delta q_{at}(t),\delta p_{at}(t),\delta q(t),\delta p(t)\right),\\ &\mathcal{N}^{T}(t)=\left(\sqrt{2\kappa}x_1^{\text{in}},\sqrt{2\kappa}y_1^{\text{in}},\sqrt{2\kappa}x_2^{\text{in}},\sqrt{2\kappa}y_2^{\text{in}},\sqrt{2f}q_{at}^{\text{in}},\sqrt{2f}p_{at}^{\text{in}},\sqrt{2\gamma_m}q^{\text{in}}, \sqrt{2\gamma_m}p^{\text{in}}\right).
\end{aligned}
\end{equation}
\end{widetext}
Our analysis requires that the system is stable, i.e., that all real parts of the eigenvalues of the drift matrix $\mathcal{A}$ are negative. These stability conditions can be derived using the Routh-Hurwitz criterion \cite{DeJesus1987}. Due to the Gaussian nature of the quantum noise and the linearity of the QLEs, the system can be fully characterized by the $8\times8$ covariance matrix (CM), which can be obtained by solving the following Lyapunov equation,
\begin{equation}
\mathcal{A}\mathcal{V}+\mathcal{V}\mathcal{A}^{T}=-\mathscr{D},
\end{equation}
where $\mathcal{V}_{ij}=\frac{1}{2}\left\{\digamma_{i}(t)\digamma_j(t^{\prime})+\digamma_j(t^{\prime})\digamma_{i}(t)\right\}$, the diffusion matrix is given by, 
\begin{equation}
\mathscr{D}=\text{diag}\left[\kappa_1,\kappa_1,\kappa_2,\kappa_2,f,f,\gamma_m(2n_{\text{th}}+1),\gamma_m(2n_{\text{th}}+1)\right],
\end{equation}
and is defined through $\langle v_{k}(t)v_{l}(t^\prime)+v_{l}(t^\prime)v_{k}(t)\rangle/2=\mathscr{D}_{kl}\delta(t-t^\prime)$.

\subsection{Quantification of bipartite and tripartite entanglement}\label{IIC}

To quantify quantum entanglement in our system, we use the logarithmic negativity $E_N$ as a measure to evaluate bipartite entanglement, while the residual minimum contangle $\mathcal{R}_{\tau}^{\min}$ is used to quantify tripartite entanglement. The logarithmic negativity is expressed as~\cite{PhysRevLett.95.119902},
\begin{equation}
E_N=\text{max}\left[0,-\text{ln}(2\varrho)\right],
\end{equation}
where $\varrho\equiv2^{-1/2}\left\{\sum(\mathcal{V})-\left[\sum(\mathcal{V})^{2}-4\text{det}(\mathcal{V})\right]^{1/2}\right\}^{1/2}$, with $\sum(\mathcal{V})=\text{det}(\psi_1)+\text{det}(\psi_2)-2\text{det}(\psi_3)$. 
The $\psi_{i}$ are the elements of the considered bipartite submatrix extracted from the covariance matrix of the whole system. For a given subsystem consisting of modes $1$ and $2$, the corresponding submatrix is,
\begin{equation}
\mathcal{V}_{sub}=
\begin{pmatrix}
\psi_1&\psi_3\\
\psi_3^{T}&\psi_2\\
\end{pmatrix}.
\end{equation}
The minimum residual contangle $\mathcal{R}_{\tau}^{\min}$ is defined as~\cite{PhysRevA.108.063704}, 
\begin{equation}
\mathcal{R}_{\tau}^{\min}\equiv min\left[\mathcal{R}_{\tau}^{c_2|ab},\mathcal{R}_{\tau}^{a|c_2b},\mathcal{R}_{\tau}^{b|c_2a}\right].
\end{equation}
This expression guarantees the invariance of tripartite entanglement under all possible permutations of the modes \cite{PhysRevA.108.063704}.  $\mathcal{R}_{\tau}^{i|jk}\equiv \mathcal{C}_{i|jk}-\mathcal{C}_{i|j}-\mathcal{C}_{i|k},~(i,j,k=c_2, a, b)$ satisfies the monogamy of quantum entanglement $\mathcal{R}_{\tau}^{i|jk}\geq 0$.
 Here, $\mathcal{C}_{u|v}$ is the contangle of the subsystem $u$ and $v$ ($v$ contains one or two modes), which can be defined as squared logarithmic negativity i.e., $\mathcal{C}_{u|v}=E^2_{\mathcal{N}_{u|v}}$, with 
\begin{equation}
E_{\mathcal{N}}\equiv \text{max}\left[0,-\ln2\eta\right],
\end{equation}
 where 
 \begin{equation}
 \eta=min~ \text{eig}[i\Omega_3\eta_{eq:6}^\prime],
 \end{equation}
  with $\Omega_3$ and $\eta_{eq:6}^\prime$ defined respectively as, \begin{equation}
  \Omega_3=\bigoplus_{k=1}^3i\sigma_{y},~~ \sigma_{y}=\begin{pmatrix}
0&-i\\i&0,
\end{pmatrix}
  \end{equation} 
  and 
  \begin{equation}
  \eta^\prime_{eq:6}=P_{i|jk}\mathcal{V}_{eq:6}P_{i|jk}~~ \text{for}~~ i,j,k=c_2,a,b,
  \end{equation}
   where 
   \begin{align}
   P_{c_2|ab}&=\text{diag(1,-1,1,1,1,1)},\\ P_{a|c_2b}&=\text{diag(1,1,1,-1,1,1)},\\ P_{b|c_2a}&=\text{diag(1,1,1,1,1,-1)},
   \end{align}
    are partial transposition matrices and $\mathcal{V}_{eq:6}$ is $6\times 6$ CM of the three modes.   
    
\subsection{Gaussian quantum discord}\label{IID}

We examine the Gaussian Quantum Discord (GQD) of the system by considering it on a pair-by-pair basis. To do this conveniently, we export a 4$\times$4 submatrix $\mathcal{V}_{sub}$ from the 8$\times$8 covariance matrix (CM), $\mathcal{V}$. For a given bipartite subsystem $j$, where $j$ is one of the 4$\times$4 matrix of the systems,  the GQD of the CM is given by  \cite{Chakraborty2017,Giorda2010},
\begin{equation}
D_G^{j}=g(\sqrt{I_1^j})-g(v^j_{-})-g(v^j_{+})g(\sqrt{\mathcal{W}^j}),
\end{equation}
where the function $g$ is defined by, 
\begin{equation}
g(x)=(x+\frac{1}{2})\ln(x+\frac{1}{2})-(x-\frac{1}{2})\ln(x-\frac{1}{2}).
\end{equation}
The simplectic eigenvalues $v^j_{-}$ and $v^j_{+}$ are given by, 
\begin{equation}
v^j_{\pm}\equiv2^{-1/2}\left\{\sum(\mathcal{V}^j_{sub})\pm\left[\sum(\mathcal{V}^j_{sub})^{2}-4I^j_{4}\right]^{1/2}\right\}^{1/2},
\end{equation}
with $\sum(\mathcal{V}_{sub})=I^j_1+I^j_2+2I^j_3$ where $I^j_1=\text{det}(\psi_1)$, $I^j_2=\text{det}(\psi_2)$, $I^j_3=\text{det}(\psi_3)$, $I^j_{4}=\text{det}(\mathcal{V}_{sub})$, and expression of $\mathcal{W}^j$ reads \cite{Chakraborty2017},
\begin{equation}
\small{\mathcal{W}^j=
\begin{cases}
\left(\frac{2|I^j_3|+\sqrt{4I_3^{j2}+(4I^j_1-1)(4I^j_{4}-I_2)}}{4I^j_1-1}\right)^2\text{if}~~\frac{4(I^j_1I^j_2-I^j_{4})^2}{(I^j_2+4I^j_{4})(1+4I^j_1)I^{j2}_3}\leq 1,\\
\frac{I^j_1I^j_2+I^j_4-I^{j2}_3-\sqrt{(I^j_1I^j_2+I^j_{4}-I^{j2}_3)^2-4I^j_1I^j_2I^j_{4}}}{2I^j_1}~~\text{otherwise}.
\end{cases}}
\end{equation}
We would like to mention that the Gaussian state is entangled when $v_{-}<1/2$. This means that if $0\leq D^j_G\leq 1$, we can have either separable or entangled states, while all states are entangled if $D^j_G>1$\cite{Giorda2010}.

\section{Results and discussion}\label{sec:III}

To carry out numerical analyses of quantum correlations in our system, we start by providing the feasible experimentally parameters used \cite{Fan2023,Zheng2024,Fan2022}:
 $\omega_m/2\pi=\SI{24}{MHz}$, $\gamma_m/2\pi=\SI{100}{Hz}$, $f/2\pi=\SI{1}{MHz}$, $\kappa_1=\SI{2}{f}$, $\kappa_2$=$\kappa_1$, $T=\SI{10}{mk}$, $\phi=\pi/2$, $\mathcal{G}_1/2\pi=\SI{2}{MHz}$, $\mathcal{G}_2/2\pi=\SI{4}{MHz}$, $J_{ac}/2\pi=\SI{12}{MHz}$, $J_{ab}/2\pi=\SI{1}{MHz}$, $\Delta^\prime_2=\Delta^\prime_1=\omega_m$ and $\Delta_{at}=-\omega_m$.

\subsection{Stability of the hybrid optomechanical system}

 \begin{figure}[htp!]
	\centering
	\includegraphics[width=.6\linewidth]{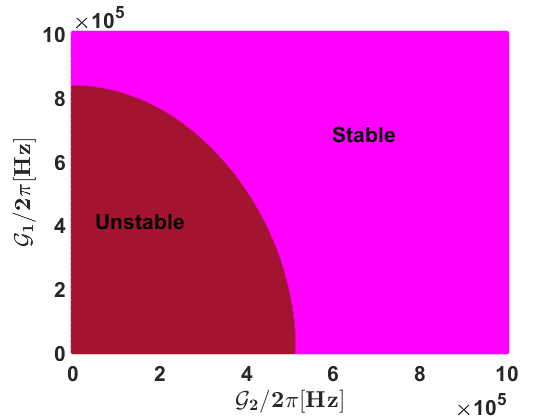}
	\caption{The dependence of the system stability on the optomechanical effective couplings $\mathcal{G}_1$ and $\mathcal{G}_2$. The stable region is marked in magenta, the unstable region in maroon. Parameters are chosen as $\omega_m/2\pi=\SI{24}{MHz}$, $\gamma_m/2\pi=\SI{100}{Hz}$, $f/2\pi=\SI{1}{MHz}$, $\kappa_1=\SI{2}{f}$, $\kappa_2$=$\kappa_1$, $T=\SI{10}{mk}$, $\phi=\pi/2$, $J_{ac}/2\pi=\SI{12}{MHz}$, $J_{ab}/2\pi=\SI{1}{MHz}$, $\Delta^\prime_2=\Delta^\prime_1=\omega_m$ and $\Delta_{at}=-\omega_m$ }
	\label{fig:2}
\end{figure}

For simplicity, we consider the cavity detuning of the two cavities to be equal, i.e. $\Delta^\prime_1=\Delta^\prime_2=\Delta$. It is worth mentioning that throughout this work we have assumed that the two cavities are resonant with the anti-Stokes sideband ($\Delta=\omega_m$) and the atoms with the Stokes sideband ($\Delta_{at}=-\omega_m$). This assumption is due to the fact that the anti-Stokes sideband can promote system cooling and facilitate the establishment of quantum entanglement, while the Stokes sideband can promote the amplification of mechanical motion, i.e., the heating of the mechanical oscillator.
 
\autoref{fig:2} illustrates the stability regions of the hybrid optomechanical system as a function of the effective optomechanical coupling strengths $\mathcal{G}_1$ and $\mathcal{G}_2$. The figure is divided into two distinct regions: a stable region (marked in magenta) and an unstable region (marked in maroon). The stability of the system is determined using the Routh-Hurwitz criterion, which ensures that the real parts of all eigenvalues of the system's drift matrix are negative. In the stable region, the system is able to maintain coherent quantum dynamics, allowing the generation and preservation of quantum correlations. This region is crucial for the practical implementation of quantum technologies, as it ensures that the system remains robust to perturbations. As the coupling strengths $\mathcal{G}_1$ and $\mathcal{G}_2$ increase, the system becomes more robust, with stability achieved for coupling strengths $\mathcal{G}_j/2\pi>\SI{1}{MHz}$ (where $j=1,2$). This indicates that sufficient optomechanical interaction is required to suppress decoherence effects caused by thermal noise and other environmental factors. In contrast, the unstable regime represents parameter values where the system becomes highly sensitive to perturbations and quantum coherence is rapidly lost. In this regime, the mechanical and optical modes are more sensitive to thermal noise, and quantum correlations cannot be maintained. The presence of instability at lower values of $\mathcal{G}_1$ and $\mathcal{G}_2$ highlights the importance of strong optomechanical coupling to achieve the desired quantum effects in the system.

The stability diagram reveals a broad parameter space where the system can support stable quantum dynamics, providing a favorable environment for generating and maintaining entanglement. As depicted in \autoref{fig:2}, the system enters an unstable regime when the optomechanical coupling is weak. In this regime, the system becomes more vulnerable to thermal noise, which disrupts quantum coherence and leads to instability.

\subsection{Generation of tripartite and bipartite entanglement in the hybrid system}

\begin{figure*}[tbh!]
	\centering
	\includegraphics[width=1\columnwidth]{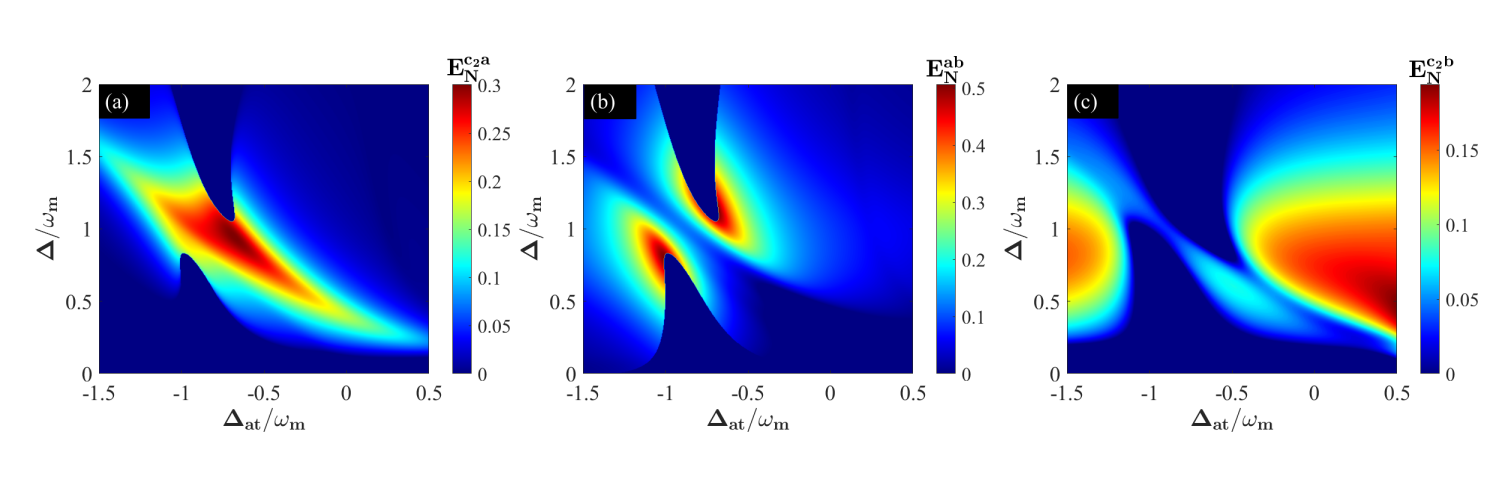}
	\caption{Contour plot of the bipartite entanglement (a) $E^{c_2a}_N$, (b) $E^{ab}_N$, and (c) $E^{c_2b}_N$ versus the normalized atomic ensemble detuning $\Delta_{at}/\omega_m$ and effective double cavity detuning $\Delta/\omega_m$. The parameters are $\omega_m/2\pi=\SI{24}{MHz}$, $\gamma_m/2\pi=\SI{100}{Hz}$, $f/2\pi=\SI{1}{MHz}$, $\kappa_1=\SI{2}{f}$, $\kappa_2$=$\kappa_1$, $T=\SI{10}{mk}$, $\phi=\pi/2$, $\mathcal{G}_1/2\pi=\SI{2}{MHz}$, $\mathcal{G}_2/2\pi=\SI{4}{MHz}$, $J_{ac}/2\pi=\SI{12}{MHz}$ and $J_{ab}/2\pi=\SI{1}{MHz}$}
	\label{fig:3}
\end{figure*}

 \begin{figure*}[tbh!]
 	\centering
 	\includegraphics[width=1\columnwidth]{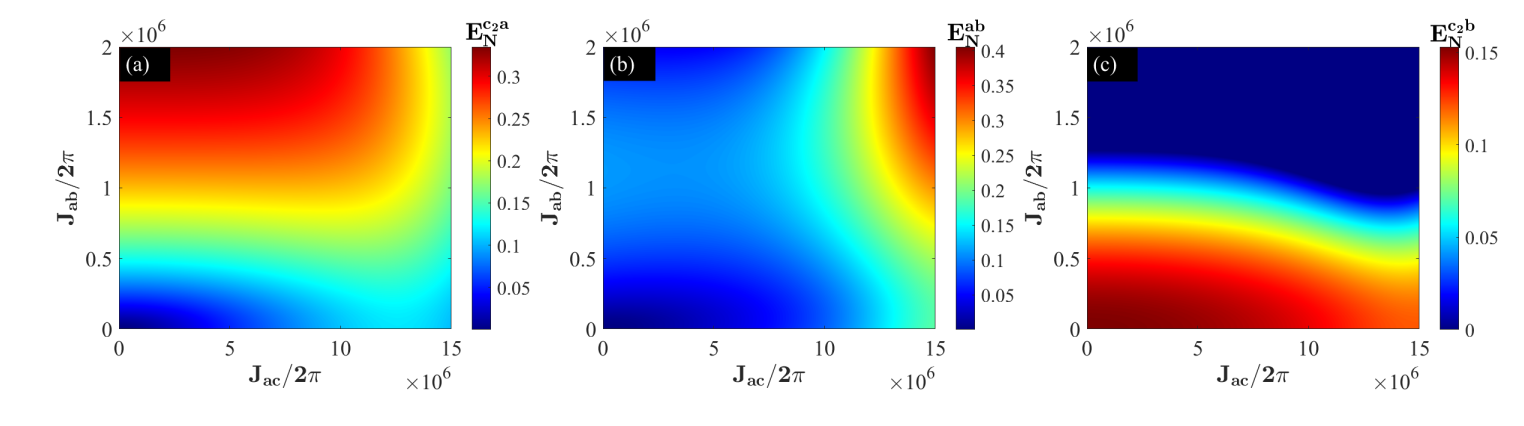}
 	\caption{Contour plot of the bipartite entanglement (a) $E^{c_2a}_N$, (b) $E^{ab}_N$, and (c) $E^{c_2b}_N$ versus the first cavity-atomic ensemble coupling constant $J_{ac}$ and mechanical oscillator-atomic ensemble coupling constant $J_{ab}$. The parameters are $\Delta^\prime_2=\Delta^\prime_1=\omega_m$ and $\Delta_{at}=-\omega_m$. The other parameters are as in \autoref{fig:3}.}
 	\label{fig:4}
 \end{figure*}

In this subsection, we explore the coexistence of both bipartite and tripartite entanglement within a hybrid optomechanical system, specifically focusing on the interactions between the following subsystems: photon mode 2, atomic ensemble, and the phonon mode. For the bipartite case, we will evaluate entanglement between photon mode 2 and the atomic ensemble $(E_N^{c_2a})$, atomic ensemble and phonon mode $(E_N^{ab})$, and the  photon mode 2 and phonon mode $(E_N^{c_2b})$. 
\begin{figure}[htp!]
\centering
\includegraphics[width=.9\linewidth]{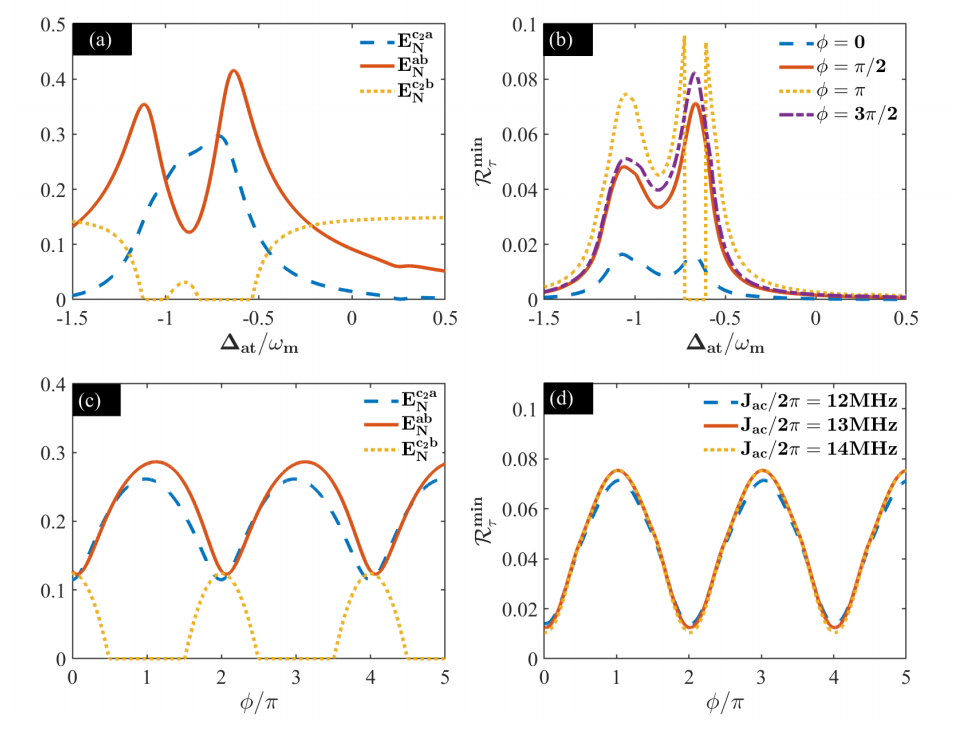}
\caption{(a) Bipartite entanglement $E^{c_2a}_N$, $E^{ab}_N$ and $E^{c_2b}_N$ versus normalized atomic ensemble detuning $\Delta_{at}$. (b) Tripartite photon-atom-phonon entanglement versus normalized atomic ensemble detuning $\Delta_{at}$ for four different values of the phase; $\phi=0$ (Azure blue--dashed line), $\phi=\pi/2$ (dark red--solid line), $\phi=\pi$ (golden yellow--dotted line) and $\phi=3\pi/2$ (purple--dash-dotted line). (c) Bipartite entanglement $E^{c_2a}_N$, $E^{ab}_N$ and $E^{c_2b}_N$ versus the phase $\phi$.  (d) Tripartite photon-atom-phonon entanglement versus the phase $\phi$ for three different values of the atomic ensemble--cavity 1 coupling constant; $J_{ac}/2\pi=12$ MHz (Azure blue--dashed line), $J_{ac}/2\pi=13$ MHz (dark red--solid line) and $J_{ac}/2\pi=14$ MHz (golden yellow--dotted line). The other parameters are as in \autoref{fig:3}}
	\label{fig:5}
\end{figure}

  \begin{figure*}[htp!]
\centering
\includegraphics[width=18cm]{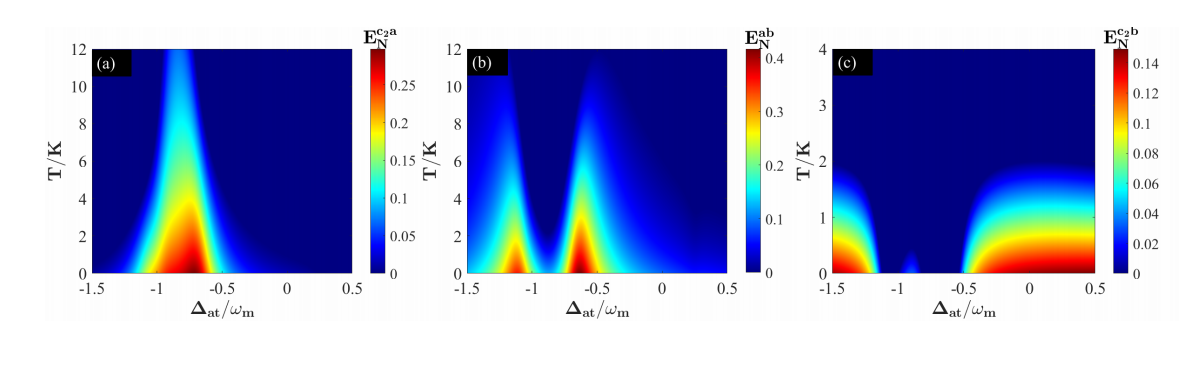}
\caption{Contour plot of bipartite entanglement $E^{c_2a}_N$, $E^{ab}_N$ and $E^{c_2b}_N$ versus the atomic ensemble detuning and the temperature $T$.  The parameters are $\Delta^\prime_2=\Delta^\prime_1=\omega_m$. The other parameters are as in \autoref{fig:3}}
\label{fig:6}
  \end{figure*}
  
\begin{figure*}[htp!]
\centering
{\includegraphics[width=19cm]{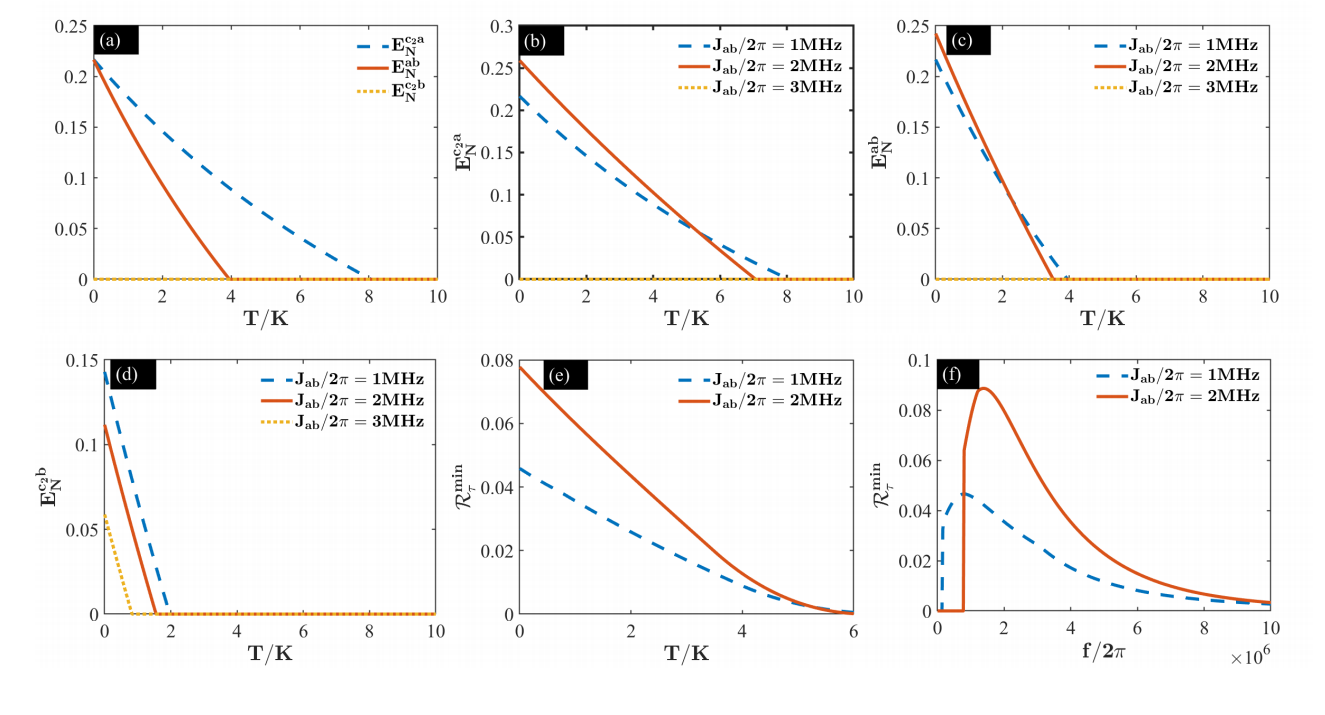}}
\caption{(a) Bipartite entanglement $E^{c_2a}_N$, $E^{ab}_N$ and $E^{c_2b}_N$ versus temperature $T$ for $\Delta_{at}=-\omega_m$. (b) Bipartite entanglement $E^{c_2a}_N$, versus temperature $T$ for different values of atomic ensemble--mechanical oscillator coupling constant; $J_{ab}/2\pi=1$ MHz (Azure blue--dashed line), $J_{ab}/2\pi=2$ MHz (dark red--solid line) and $J_{ab}/2\pi=3$ MHz (golden yellow--dotted line). The atomic ensemble detuning $\Delta_{at}=-\omega_m$. (c) Bipartite entanglement $E^{ab}_N$, versus temperature $T$ for different values of atomic ensemble-- mechanical oscillator coupling constant; $J_{ab}/2\pi=1$ MHz (Azure blue--dashed line), $J_{ab}/2\pi=2$ MHz (dark red--solid line) and $J_{ab}/2\pi=3$ MHz (golden yellow--dotted line). The atomic ensemble detuning $\Delta_{at}=-\omega_m$. (d) Bipartite entanglement $E^{c_2b}_N$, versus temperature $T$ for different values of atomic ensemble--mechanical oscillator coupling constant; $J_{ab}/2\pi=1$ MHz (Azure blue--dashed line), $J_{ab}/2\pi=2$ MHz (dark red--solid line) and $J_{ab}/2\pi=3$ MHz (golden yellow--dotted line). The atomic ensemble detuning $\Delta_{at}=-1.5\omega_m$. (e) Tripartite entanglement versus temperature $T$ for different values of the atomic ensemble-- mechanical oscillator coupling constant; $J_{ab}/2\pi=1$ MHz (Azure blue--dashed line) and $J_{ab}/2\pi=2$ MHz (dark red--solid line). The atomic ensemble detuning $\Delta_{at}=-\omega_m$. (f) Tripartite entanglement versus atomic ensemble decay rate $f$ for different values of atomic ensemble--mechanical oscillator coupling constant; $J_{ab}/2\pi=1$ MHz (Azure blue--dashed line) and $J_{ab}/2\pi=2$ MHz (dark red--solid line). The atomic ensemble detuning $\Delta_{at}=-\omega_m$. The other parameters used in \autoref{fig:7} are as in \autoref{fig:3}.}
	\label{fig:7}
\end{figure*}

When the cavities are resonant in the anti-Stokes sideband, i.e., $\Delta\approx\omega_m$, the optical drivings remove energy from the mechanical mode, thereby cooling the mechanical mode. \autoref{fig:3}(a-b) provides clear evidence that cooling the mechanical mode is a prerequisite to generate bipartite entanglement in our system. This is  achieved when the atomic frequency is around the Stokes sideband, i.e., $\Delta_{at}\approx-\omega_m$. Moreover, below the cavity resonance frequency, i.e., at $\omega_m\gg\Delta$ (off-resonance), the entanglement of the subsystems degrades to zero, i.e., 
$$E^{c_2a}_N= E^{ab}_N=E^{c_2b}_N=0,$$ 
as shown in \autoref{fig:3}. However, \autoref{fig:3}c shows that the bipartite cavity 2-mechanical mode entanglement, $E^{c_2b}_N$, is maximized for $\Delta/\omega_m>0.4$ and minimized for $\Delta/\omega_m<0.2$. This indicates that when $\Delta/\omega_m<0.2$, the degree to which energy is removed from the mechanical mode decreases significantly, leading to a very weak interaction between cavity 2 and the mechanical mode. Additionally,  this figure illustrates that the two modes are not entangled when the atomic frequency is in the Stokes sideband, i.e., $\Delta_{at}\approx-\omega_m$, where the cavity mode is at off-resonance. 

\autoref{fig:3}(b) demonstrates that maximum entanglement is reached just at the limit of zero entanglement, revealing the fragile nature of atomic ensemble-mechanical mode bipartite entanglement at this critical point (see around  $\Delta_{at}/\omega_m\approx-1$ and $\Delta/\omega_m\approx0.8$). As cavity detuning is gradually increased from  $\Delta/\omega_m\approx0.8$, the interaction between the atomic ensemble and mechanical mode strengthens, leading to an enhanced quantum correlations and maximum entanglement. A slight decrease in cavity detuning at this point weakens interactions between these modes, resulting in zero entanglement. This also indicates that the system is easily perturbed by environmental interactions, which can lead to decoherence effects. From \autoref{fig:4}(a,b), it can be observed that $E^{c_2a}_N$ enhances as the coupling rate $J_{ab}$ increases, specifically for $J_{ab}/2\pi>\SI{0.5}{MHz}$. Conversely, a different behavior occurs in \autoref{fig:4}(c), where $E^{c_2b}_N$ remains strong for weak values of $J_{ab}$. There exists a critical point above which entanglement between cavity 2 and the mechanical mode becomes zero. This phenomenon arises because as the interaction between the atomic ensemble and mechanical modes intensifies, the rate at which cavity 2-mechanical mode entanglement is partially distributed to cavity 2-atomic ensemble and atomic ensemble-mechanical mode subsystems increases, resulting in decreasing of cavity 2-mechanical mode entanglement. Furthermore, it should be noted that these analyzed photon-phonon entanglements also depend on the value of coupling rate $J_{ac}$, as shown in \autoref{fig:4}.

The effects of the atomic detuning on entanglement are analyzed in \autoref{fig:5}(a,b), while the phase $\phi$ effects are investigated in \autoref{fig:5}(c,d). Initially, bipartite entanglement is created in the optomechanical system and then partially distributed over cavity 2-atomic ensemble via atomic ensemble-mechanical oscillator interaction. This entanglement is further distributed over the atomic ensemble-mechanical mode via a direct interaction, giving rise to photon-atom-phonon genuine tripartite entanglement in our system. Notably, this genuine tripartite entanglement remains robust against the phase $\phi$ at off-resonance with respect to atomic frequency and achieves maximum for $\phi=\pi$, as shown in \autoref{fig:5}(b). As illustrated in \autoref{fig:5}(c,d), bipartite entanglement exhibits periodic behavior over the phase. The cavity 2-mechanical mode entanglement $E^{c_2b}_N$ reaches its maximum when phase is $\phi=n\pi$ (with $n$ being an even integer) and its minimum when phase is $\phi=n\pi$ (with $n$ being an odd integer). The cavity-atomic ensemble entanglement $E^{c_2a}_N$ and atomic ensemble-mechanical mode entanglement $E^{ab}_N$ are minimized when $E^{c_2b}_N$ is maximized and vice versa. \autoref{fig:5}(b,d) indicates that both bipartite and tripartite entanglements can be enhanced by tuning the phase of coupling $J_{ac}$. For $\phi=0$, photon-atom-phonon tripartite entanglement is minimal; however, maximum values are achieved at phase $\phi=\pi$. This genuine tripartite entanglement increases with enhanced coupling strength between cavity 2 and atomic ensemble (see \autoref{fig:5}(d)). In general, our results show that the entanglement in this hybrid system can be robustly controlled and enhanced by adjusting coupling parameters such as the mechanical-atomic coupling strength $J_{ab}$ and the phase $\phi$ of the system. Phase tuning and coupling strength adjustments provide efficient mechanisms for engineering and maximizing both bipartite and tripartite entanglement in the system. These findings not only demonstrate the flexibility of hybrid optomechanical systems in generating multipartite quantum correlations but also open new avenues for applications in quantum information processing.

\subsection{Temperature robustness of quantum correlations}

In this subsection, we analyze the impact of temperature on both bipartite and tripartite entanglement in our system. \autoref{fig:6} illustrates how the bipartite entanglement between the subsystems behaves as a function of atomic ensemble detuning $(\Delta_{\text{at}})$ and temperature $(T)$. The photon-atomic ensemble entanglement $(E_N^{c_2a})$ is generated around $\Delta_{\text{at}}/\omega_m \in [-1, -0.5]$ as the temperature increases, with a peak observed at approximately $\Delta_{\text{at}} \approx -0.6\omega_m$ (\autoref{fig:6}(a)).  Similarly, atomic ensemble-phonon bipartite entanglement $(E_N^{ab})$ shows a non-monotonic dependence on temperature, with two distinct peaks occurring near $\Delta_{\text{at}} \approx -0.6\omega_m$ and $\Delta_{\text{at}} \approx -1.2\omega_m$ (\autoref{fig:6}(b)). However, the photon-phonon entanglement $(E_N^{c_2b})$ exhibits a more complex behavior with three peaks, the largest of which appears around $\Delta_{\text{at}} \approx -1.5\omega_m$, as shown in \autoref{fig:6}(c).

As demonstrated in \autoref{fig:7}, the system’s sensitivity to temperature is particularly pronounced in the case of photon-phonon bipartite entanglement $(E_N^{c_2b})$, which decays rapidly as the temperature increases. In contrast, photon-atomic ensemble $(E_N^{c_2a})$ and atomic ensemble-phonon $(E_N^{ab})$ entanglements show greater resilience to thermal fluctuations, with both types of bipartite entanglement surviving up to higher temperatures (\autoref{fig:7}(a-b)). For instance, the optimal values of logarithmic negativity for photon-atomic ensemble $(E_N^{c_2a})$ and atomic ensemble-phonon $(E_N^{ab})$ bipartite entanglements are achieved at $J_{ab}/2\pi = 2 \, \text{MHz}$, with $E_N^{c_2a} \approx 0.26$ and $E_N^{ab} \approx 0.24$ (\autoref{fig:7}(a)). \autoref{fig:7}(d) reveals that at higher temperatures, the entanglement between photon mode 2 and the phonon mode $(E_N^{c_2b})$ vanishes entirely due to decoherence effects, whereas photon-atomic ensemble $(E_N^{c_2a})$ and atomic ensemble-phonon $(E_N^{ab})$ entanglements persist, showing robustness against thermal noise. This resilience is further enhanced by strong coupling between the atomic ensemble and the mechanical mode, as seen in \autoref{fig:7}(b-d). Finally, \autoref{fig:7}(e) highlights the behavior of tripartite entanglement as a function of temperature. The tripartite entanglement, quantified by the minimum residual contangle $(R_{\min}^\tau)$, remains significant even at higher temperatures, particularly when the coupling strength between the atomic ensemble and the mechanical mode $(J_{ab})$ is large. This suggests that strong atomic ensemble-mechanical mode coupling plays a crucial role in preserving tripartite entanglement at higher temperatures. The above results demonstrate that while photon-phonon entanglement $(E_N^{c_2b})$ is more vulnerable to thermal noise, photon-atomic ensemble $(E_N^{c_2a})$ and atomic ensemble-phonon $(E_N^{ab})$ bipartite entanglements, as well as tripartite entanglement, can survive at higher temperatures. The strong coupling between the atomic ensemble and the mechanical mode effectively enhances the system’s robustness to temperature-induced decoherence.

\subsection{Gaussian quantum discord}

 \begin{figure*}[htp!]
\centering
\includegraphics[width=18cm]{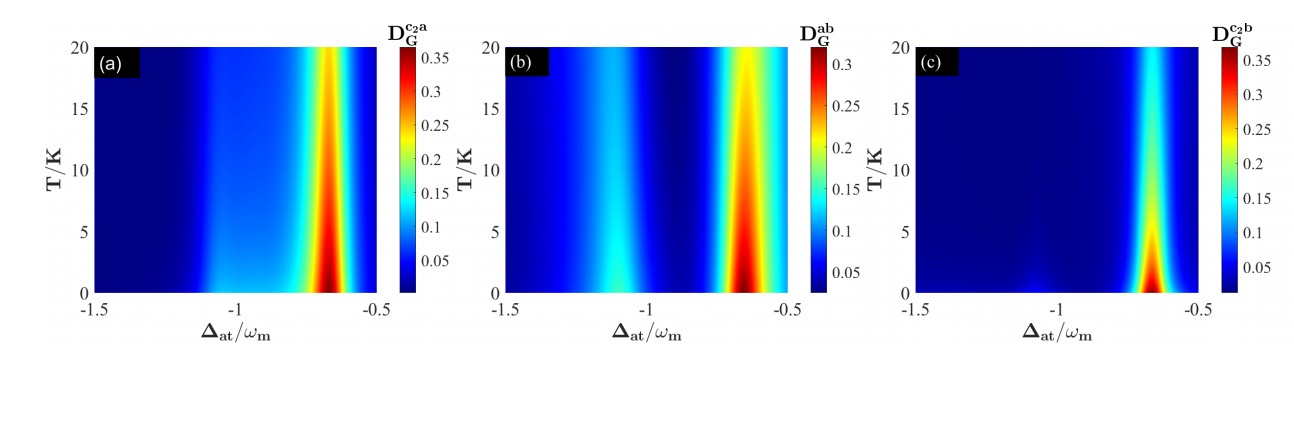}
\caption{Contour plot of GQD $D^{c_2a}_G$, $D^{ab}_G$ and $D^{c_2b}_G$ versus atomic ensemble detuning and temperature $T$. The parameters used are the same as in \autoref{fig:6}.}
\label{fig:8}
 \end{figure*} 

\begin{figure}[htp!]
\centering
\includegraphics[width=.9\linewidth]{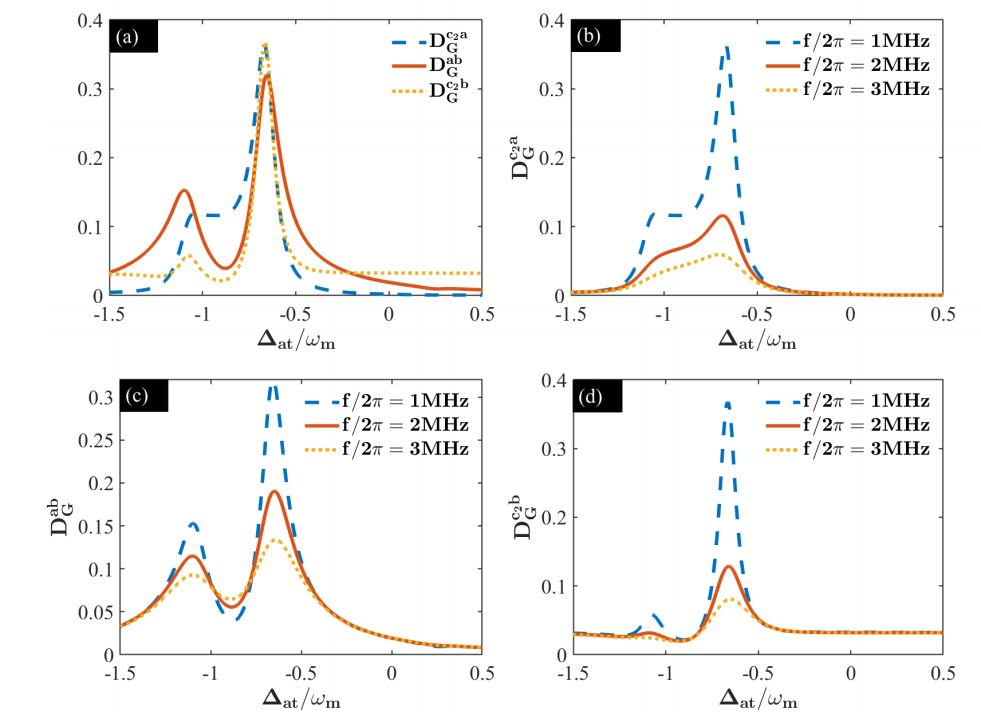}
\caption{ (a) GQD $D^{c_2a}_G$, $D^{ab}_G$ and $D^{c_2b}_G$ versus atomic ensemble detuning. (b) GQD $D^{c_2a}_G$ versus atomic ensemble detuning for different values of atomic ensemble decay rate; $f/2\pi=1$ MHz (Azure blue--dashed line), $f/2\pi=2$ MHz (dark red--solid line) and $f/2\pi=3$ MHz (golden yellow--dotted line). The cavity detuning $\Delta^\prime_2=\Delta^\prime_1=\omega_m$. (c) GQD $D^{ab}_G$ versus atomic ensemble detuning for different values of atomic ensemble decay rate; $f/2\pi=1$ MHz (Azure blue--dashed line), $f/2\pi=2$ MHz (dark red--solid line) and $f/2\pi=3$ MHz (golden yellow--dotted line). The cavity detuning $\Delta^\prime_2=\Delta^\prime_1=\omega_m$. (d) GQD $D^{c_2b}_G$ versus atomic ensemble detuning for different values of atomic ensemble decay rate; $f/2\pi=1$ MHz (Azure blue--dashed line), $f/2\pi=2$ MHz (dark red--solid line) and $f/2\pi=3$ MHz (golden yellow--deotted line). The cavity detuning $\Delta^\prime_2=\Delta^\prime_1=\omega_m$. The other parameters are as in \autoref{fig:3}}
\label{fig:9}
\end{figure}

In addition to entanglement, Gaussian quantum discord (GQD) serves as a critical measure for capturing quantum correlations that go beyond entanglement. In this subsection, we investigate the behavior of GQD in the hybrid optomechanical system. As shown in \autoref{fig:8}, we analyze GQD between the three bipartite subsystems: photon mode 2-atomic ensemble $(D_G^{c_2a})$, atomic ensemble-phonon $(D_G^{ab})$, and photon mode 2-phonon $(D_G^{c_2b})$, as functions of atomic ensemble detuning $(\Delta_{\text{at}})$ and temperature $(T)$.  \autoref{fig:8}(a) reveals that the GQD $(D_G^{c_2a})$ between photon mode 2 and the atomic ensemble shows remarkable robustness against temperature, with quantum correlations persisting even at high temperatures $(T\gg\SI{10}{K})$. The atomic ensemble-phonon GQD $(D_G^{ab})$ also demonstrates strong correlations, particularly at off-resonance detuning $(\Delta_{\text{at}} \approx -1.0\omega_m)$ (\autoref{fig:8}(b)). However, photon-phonon quantum correlations $(D_G^{c_2b})$ show a significant decay as temperature increases, with weaker correlations in comparison to the other subsystems (\autoref{fig:8}(c)). The influence of atomic ensemble detuning on GQD is further analyzed in \autoref{fig:9}.  \autoref{fig:9}(a) depicts the GQD behavior of the three subsystems versus the atomic-ensemble detuning, and it can be seen that the GQD between photon mode 2 and the atomic ensemble $(D_G^{c_2a})$ is dominant in this system, and is generated over a wide range of detuning $(\Delta_{\text{at}}/\omega_m \in [-1, -0.5])$. Moreover, \autoref{fig:9}(b-d) shows the role of atomic decay rate $f$ on GQD. As the atomic decay rate increases, quantum correlations decrease, which drop for higher atomic decay rates (\autoref{fig:9}(b,c)). This indicates that higher atomic decay rates negatively affect the quantum correlations in the system, reducing them to near-zero values for sufficiently large $f$. Furthermore, \autoref{fig:9}(d) highlights that the photon-phonon subsystem $(D_G^{c_2b})$ is the most sensitive to changes in the atomic decay rate. The rapid decay of quantum correlations in this subsystem suggests that it is the most vulnerable to decoherence, while photon-atomic ensemble and atomic ensemble-phonon correlations exhibit greater resilience.

\subsection{Phase tuning as a control mechanism for quantum correlations}

\begin{figure}[htp!]
\centering
\includegraphics[width=10cm]{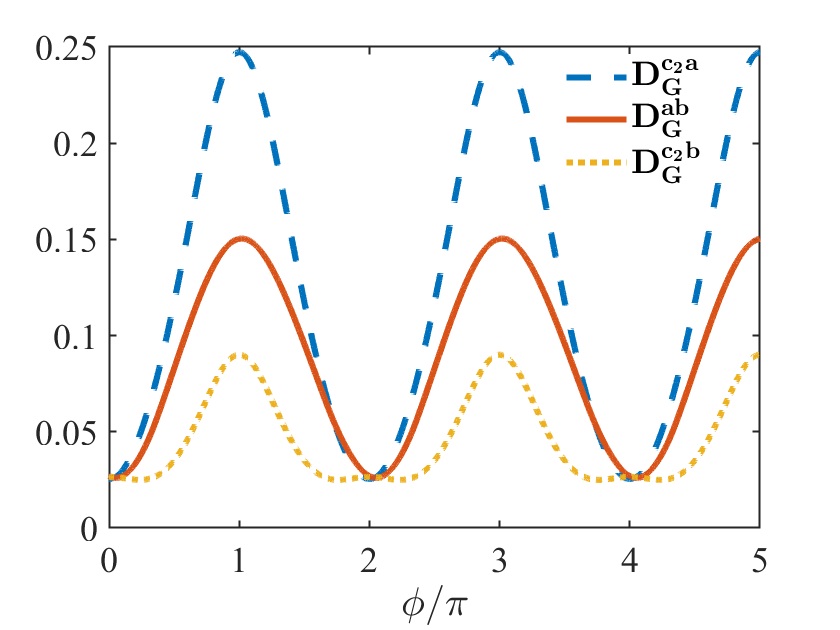}
\caption{Variation of Gaussian quantum discords, $D^{c_2a}_G$ (Azure blue--dashed line), $D^{ab}_G$ (dark red--solid line) and $D^{c_2b}_G$ (golden yellow--dotted line)  with phase $\phi$. The parameters are $\Delta^\prime_2=\Delta^\prime_1=\omega_m$ and $\Delta_{at}=-\omega_m$. The other parameters are as in \autoref{fig:3}.}
\label{fig:10}
\end{figure}

In this subsection, we investigate how phase tuning can be used as an effecient control mechanism to manipulate quantum correlations in the hybrid optomechanical system. As displayed in \autoref{fig:10}, the Gaussian quantum discord (GQD) between the three bipartite subsystems—photon mode 2-atomic ensemble $(D_G^{c_2a})$, atomic ensemble-phonon $(D_G^{ab})$, and photon mode 2-phonon $(D_G^{c_2b})$—exhibits an oscillatory behavior as a function of the phase $\phi$. This figure clearly shows that the GQD in all three subsystems reaches its maximum when the phase $\phi = n\pi$ (where $n$ is an odd integer) and its minimum when $\phi = n\pi$ (where $n$ is an even integer). This phase dependence provides a straightforward way to enhance quantum correlations by simply adjusting the phase of the coupling between the optical and atomic components of the system. Interestingly, as seen in \autoref{fig:10}, the photon mode 2-atomic ensemble correlation $(D_G^{c_2a})$ is particularly sensitive to phase tuning.
The periodic nature of GQD with phase tuning demonstrates the potential for precise control of quantum correlations in the system. By carefully selecting the phase, one can optimize the distribution of quantum correlations between the subsystems, making phase tuning a valuable tool for applications in quantum information processing and communication. Moreover, the results shown in \autoref{fig:10} suggest that phase tuning could also be used to stabilize quantum correlations in the presence of environmental noise or other disturbances. The oscillatory behavior of GQD, bounded between zero and its maximum values, suggests that certain phase settings can mitigate the degradation of quantum correlations, providing robustness against decoherence effects.

Finally, GQD provides a broader understanding of quantum correlations beyond entanglement. The results from \autoref{fig:8}, \autoref{fig:9}, and \autoref{fig:10} demonstrate that GQD remains robust at high temperatures, especially for photon-atomic ensemble and atomic ensemble-phonon subsystems. Phase-tuning provides an additional control mechanism, while the atomic decay rate poses a significant challenge to maintain strong correlations. One can conclude that, phase tuning emerges as a powerful and flexible mechanism for controlling and enhancing quantum correlations in hybrid optomechanical systems. The findings from \autoref{fig:10} highlight the importance of phase adjustments in engineering optimal quantum correlations, allowing for tailored entanglement and discord properties based on specific requirements in quantum technology applications.

\subsection{Comparison with existing approaches}

We compare here our results on quantum correlations and phase-tuning in hybrid optomechanical systems with similar approaches found in the literature. While several studies have focused on bipartite entanglement in optomechanical systems, our work extends these results by providing a comprehensive analysis of both bipartite and tripartite entanglement, as well as Gaussian quantum discord (GQD) across various subsystems.

For instance, the bipartite entanglement generation in optomechanical systems driven by squeezed light, as discussed by Bougouffa and Al-Hmoud \cite{Bougouffa2020}, primarily investigates two-mode entanglement using external squeezed states. In contrast, our system generates both bipartite and tripartite entanglement intrinsically via phase tuning without the need for external squeezed fields. \autoref{fig:3} and \autoref{fig:5} clearly demonstrate the effectiveness of phase tuning in controlling both types of entanglement, which distinguishes our approach from previous studies reliant on external driving mechanisms.

In terms of tripartite entanglement, Hmouch \textit{et al.} \cite{Hmouch2023} explored a double-cavity optomechanical system, which, like our model, included atomic ensembles interacting with mechanical oscillators. However, their approach lacked the explicit control of quantum correlations via phase tuning. Our findings, as shown in \autoref{fig:5}, reveal that the phase $\phi$ plays a critical role in optimizing tripartite entanglement, a feature not addressed in Hmouch's work. The periodic oscillation of entanglement as a function of phase (\autoref{fig:5}(b)) provides a significant advantage in dynamically controlling quantum correlations, which can be useful in quantum communication and computing applications.

Furthermore, quantum discord has gained considerable attention as a measure of quantum correlations that exist even in the absence of entanglement. Giorda and Paris \cite{Giorda2010} were among the first to establish GQD as a quantifier for continuous variable systems. However, their analysis did not explore the robustness of discord against temperature fluctuations. In contrast, our results (\autoref{fig:8}) show that GQD can survive at elevated temperatures, especially in the photon-atomic ensemble and atomic ensemble-phonon subsystems $(D_G^{c_2a}, D_G^{ab})$. This temperature robustness extends the practical applicability of GQD in real-world quantum systems, where temperature control is often challenging.

Additionally, previous studies have highlighted the challenge of maintaining quantum correlations in the presence of decoherence. Chakraborty and Das \cite{Chakraborty2023} addressed nonreciprocal photon-phonon entanglement in optomechanical systems, but their work did not focus on phase-tuning as a control mechanism. Our approach demonstrates that by carefully selecting the phase $\phi$, quantum correlations can be optimized and maintained despite decoherence, as evidenced by the results shown in Figure \ref{fig:10}. This phase-tuning strategy offers a flexible and effective method for mitigating the effects of environmental noise and improving the stability of quantum correlations.

Compared to existing approaches, our study offers several key improvements: (i) A unified framework for both bipartite and tripartite entanglement, with control via phase tuning, unlike approaches based on external squeezed fields or static systems; (ii) detailed analysis of Gaussian quantum discord, including its resilience to high temperatures, which extends beyond previous GQD studies focused solely on low-temperature regimes; (iii) the introduction of phase tuning as a robust control mechanism to dynamically optimize quantum correlations in the face of decoherence, providing practical advantages for quantum technologies. (iv) Phase tuning optimizes coupling between modes because it affects the coupling strength directly. By meticulously tuning the phase, one can maximize the interaction  between modes leading to generation of enhance entanglement. Phase tuning can reduce the system sensitivity to noise thereby making it to be more robust against the decoherence effect. Since phase tuning affects the system's dynamics directly, it can facilitate rapid responses of the system thereby enhancing entanglement on a shorter time scale. These improvements position our system as a versatile platform for exploring multipartite quantum correlations with broad applicability in quantum information processing, communication, and sensing technologies.
\subsection{Experimental realization of the phase tunin between coupling strengths}
Since the system under consideration consists of an atomic ensemble  trapped in an optical cavity, the phase of the coupling between the atomic ensemble and the cavity field can be controlled by adjusting the frequency of laser beam incident in the cavity using acousto-optic modulator. By carefully tuning the laser frequency with the acousto-optic modulator relative to atomic ensemble transitions, one can manipulate the interaction strength and phase of the coupling.
\section{conclusion}\label{sec:IV}

In this work, we have presented a comprehensive analysis of bipartite and tripartite quantum correlations, including Gaussian quantum discord (GQD), in a hybrid optomechanical system comprising two optical cavities and an atomic ensemble coupled to a mechanical oscillator. By leveraging phase tuning as a key control parameter, we have demonstrated the ability to dynamically enhance and optimize quantum correlations in the system. One of the central findings of this study is the effective use of phase tuning to control both bipartite and tripartite entanglement. As shown in \autoref{fig:5} and \autoref{fig:10}, the phase $\phi$ of the coupling between cavity 1 and the atomic ensemble significantly influences the distribution and strength of quantum correlations. This provides a flexible mechanism for fine-tuning entanglement and discord in the system, offering practical advantages for quantum technologies such as quantum communication and quantum information processing. Moreover, our study reveals that quantum correlations in the system are remarkably robust against temperature fluctuations. As demonstrated in  \autoref{fig:7} and \autoref{fig:8}, bipartite entanglement between the photon-atomic ensemble and atomic ensemble-phonon subsystems, as well as Gaussian quantum discord, can persist at higher temperatures. Our findings highlight the versatility and robustness of the hybrid optomechanical system in generating and controlling multipartite quantum correlations. The ability to dynamically adjust quantum correlations through phase tuning and to maintain these correlations in the presence of thermal noise makes this system a promising platform for future advancements in quantum information technologies. The promising results of phase-tuning and temperature robustness invite further research into other control mechanisms, such as detuning modulation or feedback control. Additionally, investigating the system's performance in realistic noisy environments will be critical for transitioning from theoretical frameworks to experimental quantum devices.

\section*{Acknowledgment}
P.D. acknowledges the Iso-Lomso Fellowship at Stellenbosch Institute for Advanced Study (STIAS), Wallenberg Research Centre at Stellenbosch University, Stellenbosch 7600, South Africa. A.K.S. acknowledges the STARS scheme, MoE, government of India (Proposal ID 2023-0161). K.S. Nisar is grateful to the funding from Prince Sattam bin Abdulaziz University, Saudi Arabia project number (PSAU/2024/R/1445). The authors are thankful to the Deanship of Graduate Studies and Scientific Research at University of Bisha for supporting this work through the Fast-Track Research Support Program.

\bibliography{Correlations}
\end{document}